\documentclass[aps,prl,twocolumn,superscriptaddress,showpacs,10pt]{revtex4-1}

\usepackage{graphicx}
\usepackage{amsmath}
\usepackage{verbatim,dsfont}
\usepackage{color}

\begin{document}
%%%%%%%%%%%%%%%%%%%%%%%%%%%%%%%%%%%%%%%%%%%%%%%%%%%%%%%%%%%%%%%%%%%%%%%%%%%%%%%%%%%%%%%%

\title{Anomalous spin-resolved point-contact transmission of holes \\ due to cubic Rashba spin-orbit coupling}

\author{Stefano Chesi}
\affiliation{Department of Physics, Purdue University, West Lafayette, IN 47907, USA}
\affiliation{Department of Physics, University of Basel, 4056 Basel, Switzerland}
\affiliation{Department of Physics, McGill University, Montreal, Quebec, Canada H3A 2T8}

\author{Gabriele F. Giuliani}
\author{L. P. Rokhinson}
\affiliation{Department of Physics, Purdue University, West Lafayette, IN 47907, USA}

\author{L.~N. Pfeiffer}
\author{K.~W. West}
\affiliation{Department of Electrical Engineering, Princeton University,
Princeton, NJ 08544 USA}

\date{\today}

%%%%%%%%%%%%%%%%%%%%%%%%%%%%%%%%%%%%%%%%%%%%%%%%%%%%%%%%%%%%%%%%%%%%%%%%%%%%%%%%%%%%%%%%%%%%%%%%%%%%%%%%%%%%%%%%%%%%%%%%%%
\begin{abstract}

Evidence is presented for the finite wave vector crossing of the two lowest one-dimensional 
spin-split subbands in quantum point contacts fabricated from two-dimensional hole gases 
with strong spin-orbit interaction. This phenomenon offers an elegant explanation for 
the anomalous sign of the spin polarization filtered by a point contact, as observed in 
magnetic focusing experiments. Anticrossing is introduced by a magnetic field parallel to 
the channel or an asymmetric potential transverse to it. Controlling the magnitude of the 
spin-splitting affords a novel mechanism for inverting the sign of the spin polarization.
\end{abstract}
%%%%%%%%%%%%%%%%%%%%%%%%%%%%%%%%%%%%%%%%%%%%%%%%%%%%%%%%%%%%%%%%%%%%%%%%%%%%%%%%%%%%%%%%%%%%%%%%%%%%%%%%%%%%%%%%%%%%%%%%%%

%72.25.-b 	Spin polarized transport
%73.23.Ad 	Ballistic transport 
%71.70.Ej 	Spin-orbit coupling, Zeeman and Stark splitting, Jahn-Teller effect 
%85.75.-d 	Magnetoelectronics; spintronics: devices exploiting spin polarized transport or integrated magnetic fields


\maketitle

\setlength\arraycolsep{2pt}

The control of spin-dependent transport in semiconductors is a central theme of fundamental and technological relevance \cite{Awschalom2002,WinklerSpringer03}. For holes, strong effects of the spin-orbit coupling have been observed in low-dimensional structures \cite{WinklerSpringer03,cubic_SOI,Rokhinson2004,Qdots_Dresselhaus,g_factor,Quay2010} and interest in the transport properties of quantum point-contacts (QPCs) has also been spurred by investigations of the so-called 0.7 anomaly \cite{Rokhinson2006,Rokhinson2008,Komijani2009}. In such hole QPCs, an intriguing and still unexplained observation is the anomalous sign of the spin polarization revealed by magnetic focusing experiments \cite{Rokhinson2004,Rokhinson2006,Rokhinson2008,Reynoso2007}.

It is well known that an asymmetric potential confining electrons or holes in 2D generates an intrinsic spin-orbit interaction (the so-called Rashba effect \cite{Bychkov1984}). However, the resulting spin-orbit coupling is very different in the two 
cases: for holes it is approximately cubic in momentum, instead of being linear 
as for electrons \cite{WinklerSpringer03,cubic_SOI}. We show here that the presence of cubic Rashba spin-orbit coupling explains the anomalous sign in the QPC transmission and, based on this, we suggest how to control the sign of the spin polarization.

%%%%%%%%%%%%%%%%%%%%%%%%%%%%%%%%%%%%%%%%%%%%%%%%%%%%%%%%%%%%%%%%%%%%%%%%%%%%%%%%%%%%%%%%%%%%%%
\begin{figure}
\includegraphics[width=0.4\textwidth]{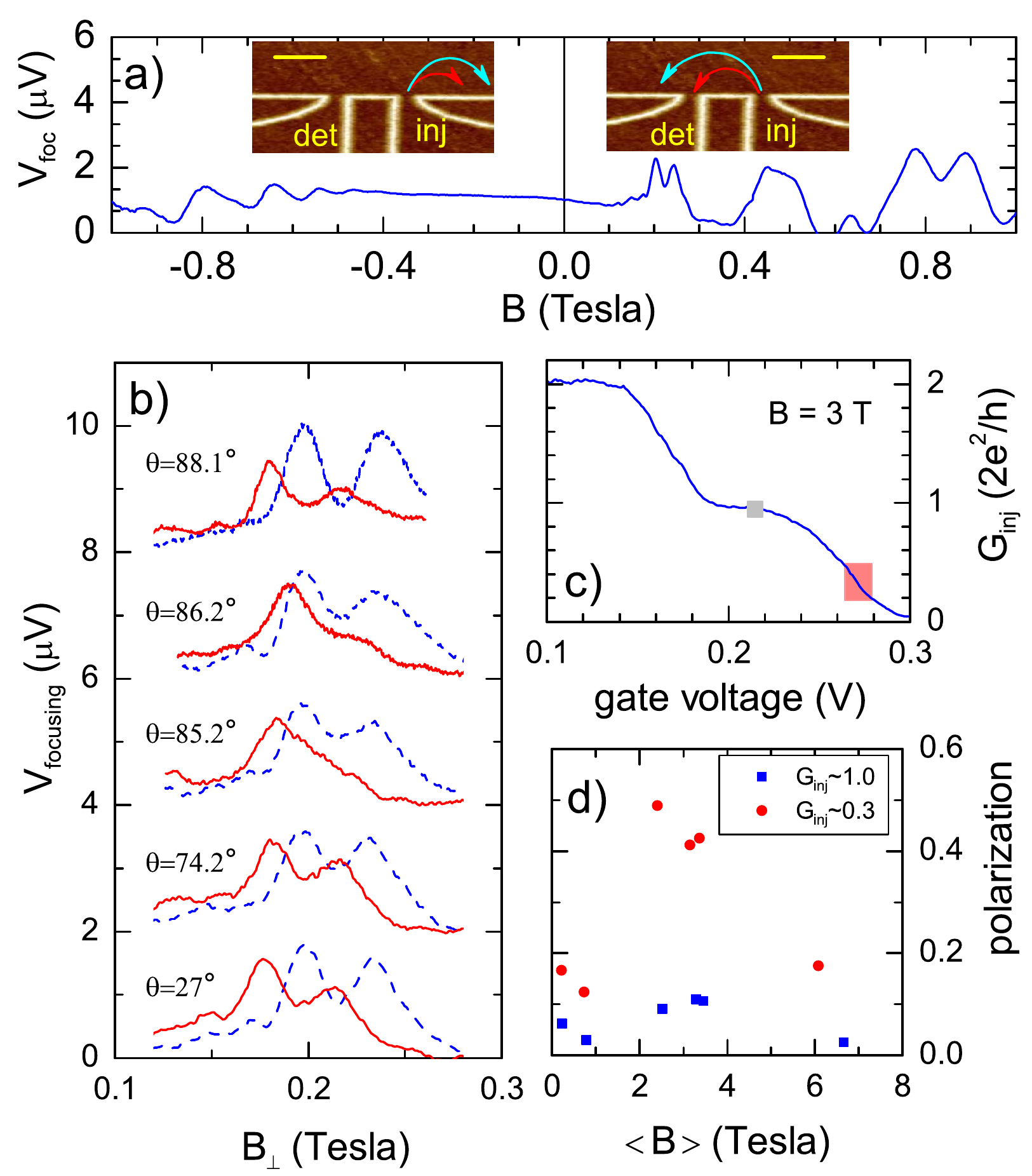}
\caption{\label{focusing} a) Voltage across the detector QPC as a function of
magnetic field for zero tilt angle. Insets: AFM micrograph of a
sample, where arrows schematically show the cyclotron motion for the two spin
orientations; the bars are 0.5 $\mu$m scales. b) Signal for the first 
focusing peak in a tilted magnetic field. Curves are offset for clarity. The
values of $G_{inj}$ for dashed blue (solid red) curves are within the smaller
(larger) rectangles in the injector QPC characteristic in c). d) Relative
population of the spin subbands, estimated for $G_{inj}=2e^2/h$ (blue squares)
and $G_{inj}\sim 0.3\cdot2e^2/h$ (red dots).}
\end{figure}
%%%%%%%%%%%%%%%%%%%%%%%%%%%%%%%%%%%%%%%%%%%%%%%%%%%%%%%%%%%%%%%%%%%%%%%%%%%%%%%%%%%%%%%%%%%%%% 

Our magnetic focusing devices are fabricated from a high mobility ($\sim 0.4
\cdot 10^6 ~{\rm V\cdot s/cm^2}$) shallow 2D hole gas \cite{Rokhinson2002}
using an atomic force microscopy (AFM) local anodic oxidation technique, see
inset in Fig.~\ref{focusing}a. The devices consist of two QPCs oriented along the $[33
\bar 2]$ crystallographic direction, with lithographical
distance $L = 0.8 ~\mu{\rm m}$ between their centers. The actual distance
is smaller due to large repulsive voltages on the side gates ($\sim0.2$ V) 
and attractive on the center gate ($-0.3$ V). Conductance of both QPCs and the nonlocal
focusing signal was measured using standard ac lock-in techniques with
excitation current 1 nA at a base temperature $T=25$ mK. The focusing signal
$V_{foc}$ is defined as the voltage across the detector QPC in response to the
current flowing through the injector QPC, see
\cite{Rokhinson2004,Rokhinson2008} for details. In the presence of
perpendicular magnetic field $B<0$, Shubnikov-de Haas (SdH) oscillations in the
adjacent 2D gas are observed, see Fig.~\ref{focusing}a, and the measured hole density
is $p = 1.45\cdot 10^{11} ~{\rm cm^{-2}}$. For $B > 0$ several peaks due to
magnetic focusing are superimposed onto the SdH oscillations. When the
conductance of both QPCs is tuned to be $2e^2/h$, the first focusing peak
splits into two peaks. If the conductance of the injector QPC is
$G_{inj}<2e^2/h$ the rightmost peak is slightly suppressed, which has been
interpreted as spontaneous polarization \cite{Rokhinson2006}.

Applying $B_{\|}$ along $[33\bar 2]$ affects the energies of the spin subbands
without affecting the cyclotron motion. Experimentally this is achieved by
tilting the sample. The focusing data in a tilted magnetic field are plotted in
Fig.~\ref{focusing}b \cite{remark_side gates}. When $G_{inj}=2e^2/h$ no filtering is expected and,
indeed, both focusing peaks have approximately the same height as at
$\theta=0$. With the increase of the tilt angle, the Zeeman
splitting of the spin subbands in a 2D gas increases. For $G_{inj}<2e^2/h$ preferential
transmission of the largest-$k_F$ spin subband is expected for electrons, which
corresponds to a suppression of the \emph{left} peak. Instead, in a hole gas we
observe suppression of the \emph{right} peak up to $\theta\approx 85^{\circ}$
($B_{\|}\approx2.5$ T), see Fig.~\ref{focusing}b. For $\theta>85^{\circ}$ the
right peak reappears. The data are summarized in Fig.~\ref{focusing}d, where
polarization $P=(V_{left}-V_{right})/(V_{left}+V_{right})$ is plotted as a function of the total field $\langle B \rangle =(B_{right} + B_{left})/2$, 
averaged between the positions of the two peaks, with $V_{left}$ and $V_{right}$ focusing
signals for the left and right peaks \cite{Rokhinson2008}.

The anomalous behavior of $P$ cannot be explained with linear Rashba spin-orbit coupling (see \cite{Reynoso2007} 
for a theoretical analysis). On the other hand, as we will show, it naturally follows from the Rashba spin-orbit coupling for 
2D holes, of the form $\frac{i\gamma}{2} \, (\hat p_-^3 \hat\sigma_+ - \hat p_+^3 \hat\sigma_-)$ 
\cite{WinklerSpringer03,cubic_SOI}. 
Here, $\hat{p}_\pm=\hat p_x\pm i \hat p_y $ and 
$\hat{\sigma}_\pm = \hat \sigma_x\pm i \hat \sigma_y $, 
with $\hat{\boldsymbol{\sigma}}$ the Pauli matrices. 
Such cubic spin-orbit interaction is responsible for a peculiar dispersion of the lowest two 
1D subbands. For a channel with lateral extent $W$, aligned with the $x$-axis, we can substitute
$\langle p_y^2 \rangle \sim \left( \hbar\pi/W \right)^2$ and $\langle p_y \rangle \sim 0$
in the 2D Hamiltonian, which gives
\begin{equation}\label{H1Dcubic}
\hat H_{1D}=\frac{\hat{p_x}^2}{2 m} + \gamma \, \left(
\frac{3\hbar^2\pi^2}{W^2} \hat p_x- \hat p_x^3 \right) 
\hat\sigma_y + 
\frac{\hbar^2 \pi^2}{2 m W^2}.
\end{equation}
Because of the lateral confinement, a linear spin-orbit coupling term appears 
in Eq.~(\ref{H1Dcubic}), which is dominant at small momenta but coexists with a cubic contribution with 
opposite sign. Therefore, the spin subbands cross not only at $k_x=0$, but also at the finite wave vectors 
$k_x = \pm \sqrt{3}\pi/W$. This is at variance with the Rashba spin-orbit splitting for electrons, 
which is monotonically increasing (linear in momentum) both in 2D and 1D. 

%%%%%%%%%%%%%%%%%%%%%%%%%%%%%%%%%%%%%%%%%%%%%%%%%%%%%%%%%%%%%%%%%%%%%%%%%%%%%%%%%%%%%%%%%%%%%%
\begin{figure}
\includegraphics[width=0.39\textwidth]{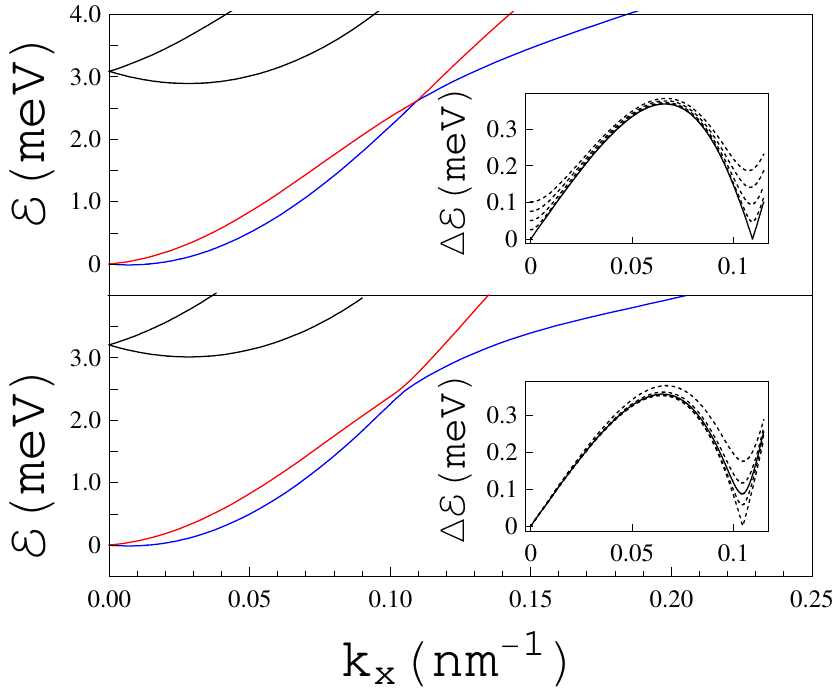}
\caption{\label{1Dsubbands} Energy subbands of 1D channels obtained from a $15~{\rm nm}$
quantum well grown in the $[113]$ direction. An electric field $\mathcal{E}_z=1~{\rm V/\mu m}$ 
along $[113]$ is present. The lateral confinement has width $W=40~{\rm nm}$. 
Upper panel: Wire along $[33\bar 2]$. The inset shows the energy splitting of the 
two lowest subbands at several values of $B_\parallel$. The solid curve is for $B_\parallel=0$ 
and the dashed curves for $B_\parallel=0.5,1,1.5,2$~T. 
Lower panel: Wire along $[1\bar{1}0]$. The inset shows the energy splitting with a lateral 
electric field. The solid curve is for ${\mathcal E}_y=0$ and the dashed curves 
for ${\mathcal E}_y=\pm 12.5, \pm 37.5~{\rm V/mm}$ (the splitting is reduced for positive 
values of ${\mathcal E}_y$).} 
\end{figure}
%%%%%%%%%%%%%%%%%%%%%%%%%%%%%%%%%%%%%%%%%%%%%%%%%%%%%%%%%%%%%%%%%%%%%%%%%%%%%%%%%%%%%%%%%%%%%%
To confirm Eq.~(\ref{H1Dcubic}), we solved the 3D problem in the framework of the 
Luttinger hamiltonian. We take into account the full cubic symmetry and 
consider a quantum well with growth direction $[113]$, as in the experiment.
%\cite{Rokhinson2004,Rokhinson2006}. 
An electric field $\mathcal{E}_z$ along the 
confinement direction produces Rashba spin-orbit coupling and the energy splitting is $\sim k^3$ in 2D. 
We then introduce a lateral confinement potential and obtain 1D subbands, plotted in Fig.~\ref{1Dsubbands}. 
For simplicity, we choose hard wall confining potentials. The 1D bands clearly display the main feature 
we are interested in: the presence of a crossing point at finite wave vector. We also checked that bulk-inversion asymmetry
terms \cite{WinklerSpringer03,RashbaSherman88} only introduce minor modifications in Fig.~\ref{1Dsubbands} and that by setting $\mathcal{E}_z=0$ a small spin splitting survives, which however does not induce crossing of the lowest two 1D subbands. For this reason, we have neglected the Dresselhaus spin-orbit terms \cite{Qdots_Dresselhaus} in the effective 2D and 1D Hamiltonians.

As seen in the inset of Fig.~\ref{1Dsubbands} (top panel), the degeneracies 
at $k_x=0$ and finite $k_x$ are removed when $B_\parallel \neq 0$. 
Within the effective Hamiltonian (\ref{H1Dcubic}), the external magnetic field is taken into 
account by adding a Zeeman term $g^* \mu_B B_\parallel \hat\sigma_x/2$, where $g^*$ is the 
effective g-factor \cite{g_factor} and $\mu_B$ the Bohr magneton. The total effective magnetic field, which 
includes the spin-orbit interaction, depends on the values of $W$ and $k_x$ as follows
\begin{equation}\label{Btot}
\vec{B}_{eff}(W,k_x)= B_\parallel \hat x + 
\frac{2\gamma \hbar^3}{g^* \mu_B}\left(\frac{3\pi^2}{W^2}k_x -k_x^3 \right) \hat y,
\end{equation}
where $\hat x,\hat y$ are unit vectors along the coordinate axes. 
The eigenstates of Eq.~(\ref{H1Dcubic}), $\psi_W(k_x,\pm) = e^{i k_x x}  |k_x,\pm \rangle_W$, 
have spinor functions $|k_x,\pm \rangle_W$ parallel/antiparallel to $\vec{B}_{eff}$ and energies
\begin{equation}\label{Einfty}
\epsilon_\pm(W,k_x)=\frac{\hbar^2 k_x^2}{2 m}\mp \frac12 g^* \mu_B |\vec{B}_{eff}(W,k_x)|~.
\end{equation}
At $k_x=0$ and $k_x = \pm \sqrt{3}\pi/W$ the spin splitting is $g^* \mu_B B_\parallel$, i.e., 
it is only due to the external magnetic field.

In a realistic QPC the width $W(x)$ of the lateral confinement changes along the channel. 
%It is large for $x \to \infty$ but a narrow constriction of width $W_0$ is present at %$x=0$. 
As in \cite{Glazman1988} we assume a sufficiently smooth variation of the width, such that the 
holes adiabatically follow the lowest \emph{orbital} subband. Introducing in Eq.~(\ref{H1Dcubic}) 
the $x$-dependent width $W(x)=W_0e^{x^2/2\Delta x^2}$, where $\Delta x$ is a typical length scale of the QPC and $W_0$ its minimum width, we obtain the following effective Hamiltonian
\begin{eqnarray}\label{PChamilt}
\frac{\hat{p_x}^2}{2 m} + V(\hat x) 
+\frac{g^*\mu_B}{2} B_\parallel\hat\sigma_x
+\gamma \, \left[3m\{V(\hat x),\hat p_x \}- \hat p_x^3 \right] \hat\sigma_y, \nonumber
\end{eqnarray}
with $\{ a, b \}= ab+ba$ \cite{comment_commutator}. The potential barrier has the following form:
\begin{equation}\label{potential}
V(x)=\frac{\hbar^2\pi^2}{2 m W(x)^2}=\frac{ \hbar^2 \pi^2}{2 m W_0^2} \, 
e^{-x^2/\Delta x^2},
\end{equation}
As it will 
be presently made clear, the main qualitative conclusions are independent of the detailed form 
of the potential, but Eq.~(\ref{potential}) allows us to solve explicitly the 1D transmission problem 
and obtain the spin-resolved conductance in the Landauer-B\"uttiker formalism. 
The scattering eigenstates are obtained with incident wavefunctions $\psi_{W=\infty}(k_\mu,\mu) $ at $x \ll -\Delta x$, where 
$\mu=\pm$ denotes the spin subband and $k_\pm$ are determined by the Fermi energy 
$\epsilon_F$, at which the holes are injected in the QPC. For $x \gg \Delta x$, such 
QPC wavefunctions have the asymptotic form $\sum_{\nu=\pm}t_{\mu,\nu}\psi_{\infty}(k_\nu,\nu) $, where $t_{\mu,\nu}$ are
transmission amplitudes. The spin-resolved conductances 
are simply given by $G_\pm=\frac{e^2}{h}\sum_{\mu=\pm}\frac{ v_\pm}{v_\mu}|t_{\mu,\pm}|^2$  
\cite{comment_unpol_incident}, where the Fermi velocities are
$v_\pm = \frac{\partial \epsilon_\pm (\infty,k_\pm)}{\partial\hbar k_x}$, 
from Eq.~(\ref{Einfty}). The total conductance is $G=G_++G_-$.
Typical results at several values of $B_\parallel$ are shown in 
Fig.~\ref{PC_transmission_fig}. As usual, by opening the QPC, a current starts 
to flow above a minimum value of $W_0$. The spin polarization behaves as follows:

%%%%%%%%%%%%%%%%%%%%%%%%%%%%%%%%%%%%%%%%%%%%%%%%%%%%%%%%%%%%%%%%%%%%%%%%%%%%%%%%%%%%%%%%%%%%%%
\begin{figure}
\begin{center}
\includegraphics[width=0.4\textwidth]{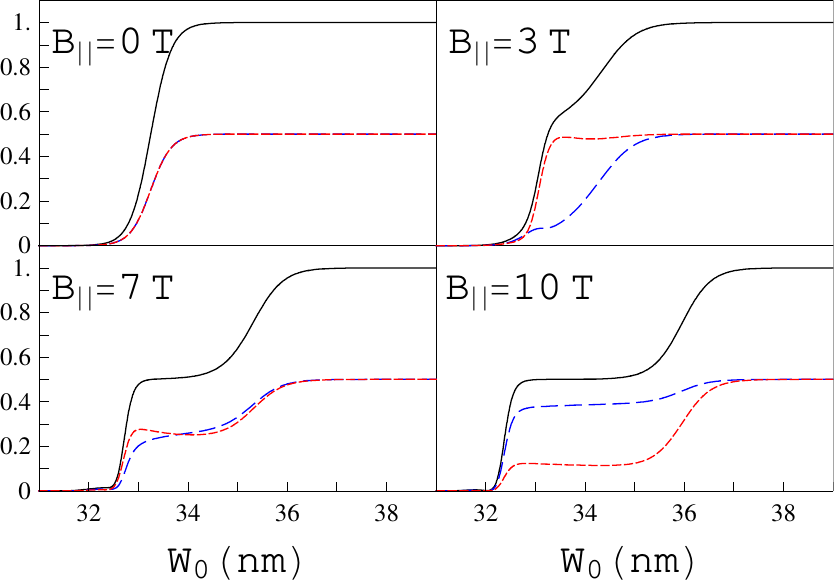}
\caption{\label{PC_transmission_fig} 
Total conductance $G$ (black solid curves) and spin-resolved conductances $G_+$ (blue, long-dashed curves) and $G_-$ (red, short-dashed curves), plotted in units of $2e^2/h$ as functions 
of the minimum width $W_0$ of the QPC [see Eq.~(\ref{potential})]. We used parameters appropriate for the experimental 
setup:  $m=0.14 m_0$ \cite{mass}, where $m_0$ is the bare electron mass, 
 $g^*=0.8$ \cite{g_factor}, $\gamma\hbar^3=0.45\,{\rm eV} \,{\rm nm}^3$, 
$\Delta x=0.3 \, \mu{\rm m}$, and  $\epsilon_F=2.3$~meV. }
\end{center}
\end{figure}
%%%%%%%%%%%%%%%%%%%%%%%%%%%%%%%%%%%%%%%%%%%%%%%%%%%%%%%%%%%%%%%%%%%%%%%%%%%%%%%%%%%%%%%%%%%%%% 

(i) At $B_\parallel = 0$ T we obtain a structureless unpolarized conductance ($G_+=G_-$) but we find $G_- > G_+$ at larger values of the magnetic field (see the top right panel of 
Fig.~\ref{PC_transmission_fig}, at $B_\|=3~{\rm T}$), i.e., the holes in the \emph{higher} 
spin subband have larger transmission at the first plateau. The sign is opposite to the case 
of linear Rashba spin-orbit coupling (see \cite{Reynoso2007}) and in agreement with the experimental 
results of Fig.~\ref{focusing}.

(ii) At $B_\parallel \approx 7$ T (see the bottom left panel of 
Fig.~\ref{PC_transmission_fig}) $G_+ \simeq G_-$ and the transmission becomes unpolarized, as 
observed in the data of Fig.~\ref{focusing}.

(iii) At even larger values of $B_\|>7$ T, we obtain $G_+ \simeq e^2/h$, $G_-\simeq 0$ 
(bottom right panel of Fig.~\ref{PC_transmission_fig}). Although this regime is yet to be 
experimentally investigated, this represents a natural prediction of our theory: at 
sufficiently large magnetic field the role of the spin-orbit coupling becomes negligible and 
the spin direction (parallel/antiparallel to the external magnetic field) of the holes is conserved. 
The injected holes remain in the original (``$+$" or ``$-$") branch and the current at the first 
plateau is polarized in the ``$+$" band, which has lower energy. Deviations from this behavior 
are due to non-adiabatic transmission in the spin subbands and, to gain a qualitative 
understanding, we consider next a semiclassical picture of the holes. 

When a hole wave-packet is at position $x$, it is subject to a magnetic field $\vec{B}_{eff}$ 
determined by $W(x)$ and $k_x(x)$ as in Eq.~(\ref{Btot}). For holes injected at $\epsilon_F$, the momentum 
is determined by energy conservation. Treating the spin-orbit coupling as a small perturbation compared to the 
kinetic energy, we have $k_x(x) \simeq \sqrt{k_F^2- \pi^2/W(x)^2}$,
where $k_F=\sqrt{2 m \epsilon_F}/\hbar$ is the Fermi wave-vector in the absence of 
spin-orbit coupling. Therefore, the injected hole experiences a varying magnetic field in its semiclassical motion along $x$, due to the change of both $k_x$ and $W(x)$. 
For adiabatic transmission of the spin subbands the spin follows the direction of the magnetic field, but this is not possible in general if 
$B_\parallel$ is sufficiently small. In particular, for $B_\parallel=0$ Eq.~(\ref{H1Dcubic}) implies 
that $\hat\sigma_y$ is conserved. Therefore, the initial spin orientation along $y$ is not affected by 
the motion of the hole. On the other hand, $\vec{B}_{eff}$ of Eq.~(\ref{Btot}) changes direction 
when $k_x= \sqrt{3} \pi/W $ and $B_\parallel =0$. After this point, a hole in the ``$+$" branch continues its motion in 
the ``$-$" branch and vice-versa. 

At finite in-plane magnetic field the degeneracy of the spectrum is removed but the holes do not follow 
adiabatically the spin branch, unless the Landau-Zener condition $\frac{dB_y/dt}{B_\parallel} \ll \omega_B$ 
is satisfied, where $\hbar\omega_B=g^* \mu_B B_\parallel$. The change $\Delta B_y$ in the spin-orbit 
field is obtained from Eq.~(\ref{Btot}): $|B_y|$ is equal to $2 \gamma \hbar^3 k_F^3/g^* \mu_B$ far from 
the QPC and vanishes at the degeneracy point. This change occurs on the length scale $\Delta x$ of the 
QPC and we can estimate the time interval with $\Delta t \simeq \Delta x/v$ where $v$ is a typical 
velocity of the hole. 
This gives  
\begin{equation}\label{LandauZener}
B_\parallel \gg \sqrt{\frac{\hbar \Delta B_y}{g^* \mu_B \Delta t}}\simeq \frac{\hbar^2 \sqrt{2\gamma k_F^3 v/\Delta x}}{g^* \mu_B}.
\end{equation} 
To estimate $v$ at the degeneracy point $k_x=\sqrt{3}\pi/W$, we solve $\sqrt{3}\pi/W \simeq \sqrt{k_F^2-\pi^2/W^2}$ to obtain $k_x = \frac{\sqrt{3}}{2}k_F$. Therefore, $v$ is large at the degeneracy point ($v\simeq v_F$, where $v_F=\hbar k_F/m$ is the Fermi velocity), and to follow adiabatically the spin branches requires a large external field. 
The crossover occurs for
\begin{equation}\label{Bcross}
B^* \simeq \frac{(\hbar k_F)^2 \sqrt{2\gamma \hbar /(m \Delta x)}}{g^* \mu_B}.
\end{equation} 
This expression gives  $B^* \simeq 7.4$ T  with the parameters of Fig.~\ref{PC_transmission_fig}, 
in agreement with the more accurate numerical analysis. 
Below $B^*$, holes injected in the ``$+$" band cross non-adiabatically to the ``$-$" spin branch 
when $k_x \simeq \sqrt{3}\pi/W$. Therefore, holes injected in the lower subband have higher energy 
at $x\simeq 0$ and are preferentially reflected, as seen in the top right panel of Fig.~\ref{PC_transmission_fig}. 
The reflection is not perfect, due to non-adiabaticity at $k_x \simeq 0$: at this second quasi-degenerate point the 
``$-$" holes can cross back to the ``$+$" branch, and be transmitted. We attribute to this effect the enhanced 
conductivity $G>e^2/h$ at the first conductance plateau in the top right panel of Fig.~\ref{PC_transmission_fig}, while a well-defined $e^2/h$ plateau is obtained at larger magnetic field. In fact, the adiabatic approximation becomes accurate at $k_x \simeq 0$ for smaller values of $B_\parallel$ \cite{comment_zero_kx} than $B^*$. 

The above discussion makes it clear that the degeneracy of the hole spectrum at $k_x=\sqrt{3}\pi/W$ 
is crucial to obtain the anomalous transmission of Figs.~\ref{focusing} and \ref{PC_transmission_fig}. The special 
behavior we have described cannot be realized with linear Rashba spin-orbit coupling \cite{Reynoso2007}.  Furthermore, Eq.~(\ref{Bcross}) allows us to predict how the value 
of the crossover field can be controlled. A lower value of $B^*$ can be obtained with a smaller 
coupling $\gamma$, a smoother QPC (i.e., larger $\Delta x$), or a lower hole gas density 
(i.e., smaller $k_F$). The value of the Fermi wave vector has a large influence, since it 
contributes both to the spin-splitting $\gamma \hbar^3 k_F^3$ and to the velocity $v_F$ of the holes. 

It is also remarkable that the degeneracy of the 1D spectrum at finite $k_x$ is removed for 
a channel oriented along the $[1\overline 1 0]$ direction, as shown in the bottom panel of 
Fig.~\ref{1Dsubbands}. The reason is that the lateral confinement is along the low symmetry 
direction $[33\overline 2]$ and the mirror symmetry of the channel is broken by the crystalline 
potential. At the anti-crossing, we obtain a $\sim 0.1$~meV splitting (see inset). For the other 
orientation of the wire this splitting corresponds to a magnetic field $B_\parallel \sim 1$~T, and it is therefore 
quite sizable. This also suggests that it should be possible to modify the spin splitting, 
and thus the crossover field $B^*$, via electric gates. We consider in the second inset of 
Fig.~\ref{1Dsubbands} an electric field $\mathcal{E}_y$ in the transverse direction of the channel 
and obtain that the splitting can be either reduced or increased by varying $\mathcal{E}_y$. 
In contrast to the case $B_\parallel \neq 0 $, the degeneracy at $k_x=0$ is not lifted 
by the transverse electric field.

In conclusion, we have shown that the cubic Rashba spin-orbit coupling for holes provides an explanation
of the anomalous sign of the spin polarization observed in QPC's in 2D hole gases. The theory nicely explains the presence 
of a crossover field $B^*$ at which the transmission is unpolarized, predicts that above $B^*$ a polarization 
in the lowest spin subband is recovered, and indicates how the value of $B^*$ 
can be modified.

We would like to thank Yu. Lyanda-Geller and E. I. Rashba for useful discussions. SC acknowledges support by
NCCR Nanoscience, Swiss NSF, and CIFAR.

%%%%%%%%%%%%%%%%%%%%%%%%%%%%%%%%%%%%%%%%%%%%%%%%%%%%%%%%%%%%%%%%%%%%%%%%%%%%%%%%%%%%%%%%%%%%%%%%%%%%%%%%%

%%%%%%%%%%%%%%%%%%%%%%%%%%%%%%%%%%%%%%%%%%%%%%%%%%%%%%%%%%%%%%%%%%%%%%%%%%%%%%%%%%%%%%%%%%%%%%%%%%%

%%%%%%%%%%%%%%%%%%%%%%%%%%%%%%%%%%%%%%%%%%%%%%%%%%%%%%%%%%%%%%%%%%%%%%%%%%%%%%%%%%%%%%%%%%%%%%%%%%

\begin{thebibliography}{99}
\bibitem{remark_side gates}
The shift seen in Fig.~\ref{focusing}b between traces at the same tilt angle  can be 
attributed to a local decrease of hole density from the side gates (0.22 V for the blue dashed traces
and 0.265-0.28 V for the red solid traces).
\bibitem{comment_commutator}
The anticommutator is introduced to obtain a Hermitian 
Hamiltonian. This has negligible effect in the limit of a smooth 
contact, when $\partial V/\partial x$ is small.
\bibitem{comment_unpol_incident}
We define $G_\pm$ with unpolarized incident holes and spin-resolved detection. The conductances 
for spin-polarized holes and unpolarized detection have the same values of $G_\pm$ in our model.
\end{thebibliography}
\end{document}